# Spatial-temporal V-Net for automatic segmentation and quantification of right ventricle on gated myocardial perfusion SPECT images

Chen Zhao[1#], Shi Shi[2#], Zhuo He[1], Saurabh Malhotra[3,4], Cheng Wang[5], Zhongqiang Zhao[5], Xinli Li[5], Haixing Wen[6], Shaojie Tang[6], Yanli Zhou[5]*, Weihua Zhou[1,7]*

1. Department of Applied Computing, Michigan Technological University, Houghton, MI, 49931, USA

2. Department of Cardiology, Hai'an People's Hospital, Nantong, 226600, China

3. Division of Cardiology, Rush Medical College, Chicago, IL, 60612, USA

4. Division of Cardiology, Cook County Health, Chicago, IL, 60130, USA

5. Department of Cardiology, The First Affiliated Hospital of Nanjing Medical University, Nanjing, China

6. School of Automation, Xi'an University of Posts and Telecommunications, Xi'an, Shaanxi 710121, China

7. Center for Biocomputing and Digital Health, Institute of Computing and Cybersystems, and Health Research Institute, Michigan Technological University, Houghton, MI, 49931, USA

**Running Title:** RV segmentation on SPECT MPI

Chen Zhao and Shi Shi contributed equally to this work.

Corresponding authors:

Weihua Zhou, PhD          E-Mail: whzhou@mtu.edu

Department of Applied Computing, Michigan Technological University

1400 Townsend Dr, Houghton, MI, 49931, USA

Or

Yanli Zhou, MD, PhD          E-mail: zhyl88@qq.com

Department of Cardiology, The First Affiliated Hospital of Nanjing Medical University,

300 Guangzhou Rd, Gulou, Nanjing, 210000, China

## Abstract

**Background.** Functional assessment of right ventricle (RV) using gated myocardial perfusion single-photon emission computed tomography (MPS) heavily relies on the precise extraction of right ventricular contours. In this paper, we present a new deep-learning-based model integrating both the spatial and temporal features in gated MPS images to perform the segmentation of the RV epicardium and endocardium.

**Methods.** By integrating the spatial features from each cardiac frame of the gated MPS and the temporal features from the sequential cardiac frames of the gated MPS, we developed a Spatial-Temporal V-Net (ST-VNet) for automatic extraction of RV endocardial and epicardial contours. In the ST-VNet, a V-Net is employed to hierarchically extract spatial features, and convolutional long-term short-term memory (ConvLSTM) units are added to the skip-connection pathway to extract the temporal features. The input of the ST-VNet is ECG-gated sequential frames of the MPS images and the output is the probability map of the epicardial or endocardial masks. A Dice similarity coefficient (DSC) loss which penalizes the discrepancy between the model prediction and the ground truth was adopted to optimize the segmentation model.

**Results.** Our segmentation model was trained and validated on a retrospective dataset with 45 subjects, and the cardiac cycle of each subject was divided into 8 gates. The proposed ST-VNet achieved a DSC of 0.8914 and 0.8157 for the RV epicardium and endocardium segmentation, respectively. The mean absolute error, the mean squared error, and the Pearson correlation coefficient of the RV ejection fraction (RVEF) between the ground truth and the model prediction were 0.0609, 0.0830, and 0.6985.

**Conclusion.** Our proposed ST-VNet is an effective model for RV segmentation. It has great promise for clinical use in RV functional assessment.

**Keywords:** Myocardial perfusion SPECT, myocardial perfusion imaging, right ventricle, deep learning, image segmentation

## 1. Introduction

Currently, myocardial perfusion single-photon emission computed tomography ([SPECT], MPS) is mainly used to detect myocardial perfusion defects and evaluate left ventricular function (LV) [1]. Studies have quantitatively evaluated right-ventricular (RV) function in MPS and demonstrated their diagnostic and prognostic values [2–5]. Nevertheless, RV functional assessment is often ignored during the MPS examination due to the challenges in identifying RV and extracting its contours in these images. It is important to develop algorithms and techniques for automated RV segmentation and subsequent functional assessment in MPS.

Most of the automated myocardial segmentation methods using MPS are based on traditional image processing and the segmentation targets is the LV. The commercial method presented by Germano et al counted the profiles for myocardial walls and employed the Gaussian filter to smoothen the boundaries [6]. Huang et al. presented a region-based active contour model to perform the LV myocardium segmentation according to the scalable fitting energy and local image fitting energy, and it achieved a Dice similarity coefficient (DSC) of 0.7946 for LV segmentation using short-axis gated MPS [7]. Soneson et al. employed the Dijkstra algorithm to find an optimal mid-mural line within the myocardial wall and then to search for the epicardial and endocardial contours based on the signal intensity and the myocardial wall thickness [8]. However, without image feature extraction and feature selection, the LV myocardium would be over- or under-segmented using traditional image processing-based methods.

In recent years, machine learning and deep learning-based methods were integrated into automatic myocardium segmentation. Betancur et al. employed a support vector machine with hand-crafted image features to perform the mitral valve plane localization, which is the prepositive step for LV myocardium segmentation and MPI quantification [9]. Wang et al. proposed a multi-channel 3D V-Net for LV segmentation in MPS [10]. Using physician-confirmed LV contours as the learning-based labels, the 3D multi-class V-Net architecture enabled voxel-wise error back-propagation during the training stage, directly outputting an equal-sized prediction patch with the input patch during the testing procedure and achieved DSCs of 0.965 and 0.910 for LV epicardial and endocardial segmentation. Wen et al. further improved V-Net and applied the implemented model to LV myocardium segmentation on gated MPS [11], and it achieved a DSC of 0.9222 for endocardium, 0.9580 for myocardium, and 0.9748 for epicardium. Due to the data-driven and automatic feature extraction, these deep learning-based methods significantly outperformed the traditional image processing methods and machine learning approaches. However, the V-Net can only extract spatial features, and the temporal features from the contextual cardiac frames in gated MPS are ignored, though they are important for sequential MPS image feature extraction.

Manual segmentation of the right ventricle (RV) is time-consuming, tedious, and subjective; thus, automated methods are desired in clinical practice. However, automatic identification of RV is challenging due to the ambiguous boundary, irregular cavity and the shape changes considerably. Evaluation of RV function is more difficult than LV because of RV anatomy [12]. The RV is pyramidal in shape, with an inlet, a main body, and an outflow tract; thus, the borders are difficult to outline [13]. In addition, RV wall thickness typically is a third of that of the LV, exacerbating difficulties of RV segmentation. The RV in general has low counts on MPS. All these factors make it much more difficult to evaluate RV anatomy and function than it is to evaluate the same for LV in MPS. Nevertheless, the RV is visible on MPS in certain patient populations including dilated cardiomyopathy and pulmonary hypertension. In our study [5], RV dyssynchrony was measured by phase analysis of gated FDG-PET in patients with pulmonary hypertension and validated against the contraction delay measured by speckle tracking echocardiography; furthermore, it showed a significant correlation with RV dysfunction. In our other study [14], both LV and RV dyssynchrony was measured from gated MPS and the difference between them was used to define the interventricular mechanical dyssynchrony; there was an excellent agreement between the interventricular mechanical dyssynchrony and bundle branch block patterns identified by ECG. Both studies employed a template-based method to sample RV myocardial walls for functional assessment, which is a major technical limitation [13]. As a result, it is important to develop a more accurate and robust segmentation method for improved RV functional assessment.

To the best of our knowledge, this paper presents the first study for RV segmentation using MPS and a deep learning method. In this paper, a deep learning-based method is presented for RV epicardium and endocardium segmentation. In detail, a spatial-temporal V-Net (ST-VNet) is presented to extract both spatial and temporal features from the 4D MPS imaging sequence. The input of the ST-VNet is a sequence of the MPS images containing the cropped volumes from gated cardiac frames, and the output of the ST-VNet is the probability map of the epicardial or endocardial masks for the volume from the last cardiac frame in the input sequence. A DSC loss, which penalizes the discrepancy between the model-predicted RV masks and the ground truth, is adopted to optimize the model. The proposed ST-VNet achieved a DSC of 0.8914 and 0.8157, and an average surface distance of 1.3479 and 2.5411 pixels for RV epicardium and endocardium segmentation, respectively.

## 2. Materials and Methods

### 2.1. Study Population

This retrospective study enrolled 45 subjects who underwent standard 12-lead ECG gated MPS from May 2012 to September 2016. All the patients had mycardial ischemia(14), or myocardial infarction(6), or heart failure(35). The average age was 58±11 years, and 61% were men with congestive heart failure. Thirty patients (66.7%) had abnormal LV myocardial perfusion on MPS. All subjects were enrolled from the First Affiliated Hospital of Nanjing Medical University. The study was approved by the Institutional Ethical Committee of the First Affiliated Hospital of Nanjing Medical University, and written informed consent forms were obtained from all participants.

### 2.2 Image Acquisition

MPS images were acquired with a dual-headed gamma camera (Philips Medical Systems, Milpitas, CA, USA) using a 2-day stress/rest protocol with $99^m$Tc-sestamibi as the radiopharmaceutical. For either stress or rest MPI, $99^m$Tc-sestamibi doses ranged from 25 to 30 mCi according to the body mass index(BMI). The imaging parameters were 20% energy window around 140KeV, 180º orbit, and 32 steps with 25 seconds per step, 8-bin gating, and 64 projections per gate. The total acquisition time was 14 minutes for each patient. Image reconstruction and reorientation were completed with Emory Reconstruction Toolbox (ERToolbox; Atlanta, GA). MPS data was reconstructed by ordered subset expectation maximization (OSEM) with 3 iterations and 10 subsets and then filtered by a Butterworth low-pass filter with a cut-off frequency of 0.4 cycles/cm and an order of 10.

Each 3D MPS image volume was first cut longitudinally to generate 20 2D long-axis slices [15], and then each slice was automatically cropped into 64×64 pixels. As a result, each scanned gate (cardiac frame) contains a 3D volume with a size of 64×64×20 and the voxel spacing is fixed at 6.4 mm. A well-trained nuclear cardiologist (S.S.) manually annotated the contours for both epicardial and endocardial myocardium for each gate of the MPS and results were confirmed by a senior nuclear cardiologist (Y.Z.). The data with manual RV annotations are shared on GitHub https://github.com/MIILab-MTU/RV_segmentation.

### 2.3. ST-VNet

We formulate RV segmentation as a spatiotemporal sequence segmentation problem in which the input is a spatiotemporal sequence and the prediction target is a 3D volume where 1 represents the epicardium or endocardium, and 0 indicates the background. Suppose that the gated MPS is denoted as $\mathcal{X}_i \in R^{H \times W \times D}$, where $i$ is the index of the gate (cardiac frame) in a gated MPS sequence, and $H$, $W$ and $D$ indicate the height, width, and depth of the MPS volume at gate $i$. The RV segmentation problem is further converted into predicting the contours of epicardium or endocardium for MPS at gate $t$, as shown in Eq. 1.

$$\mathcal{Y}_{i+t} = \underset{\mathcal{X}_i,\dots,\mathcal{X}_{i+t}}{\mathrm{argmax}}\, p(Y_{i+t}|\mathcal{X}_i, \mathcal{X}_{i+1}, \dots, \mathcal{X}_{i+t}) \tag{1}$$

where $\mathcal{Y}_i \in R^{H \times W \times D}$ is the mask of the epicardium or endocardium of the MPS $\mathcal{X}_i$.

The proposed MPS RV myocardium segmentation model, which is called ST-VNet, contains a V-Net and convolutional long-term short-term memory (LSTM) units. The architecture of the proposed ST-VNet is shown in Figure 1.

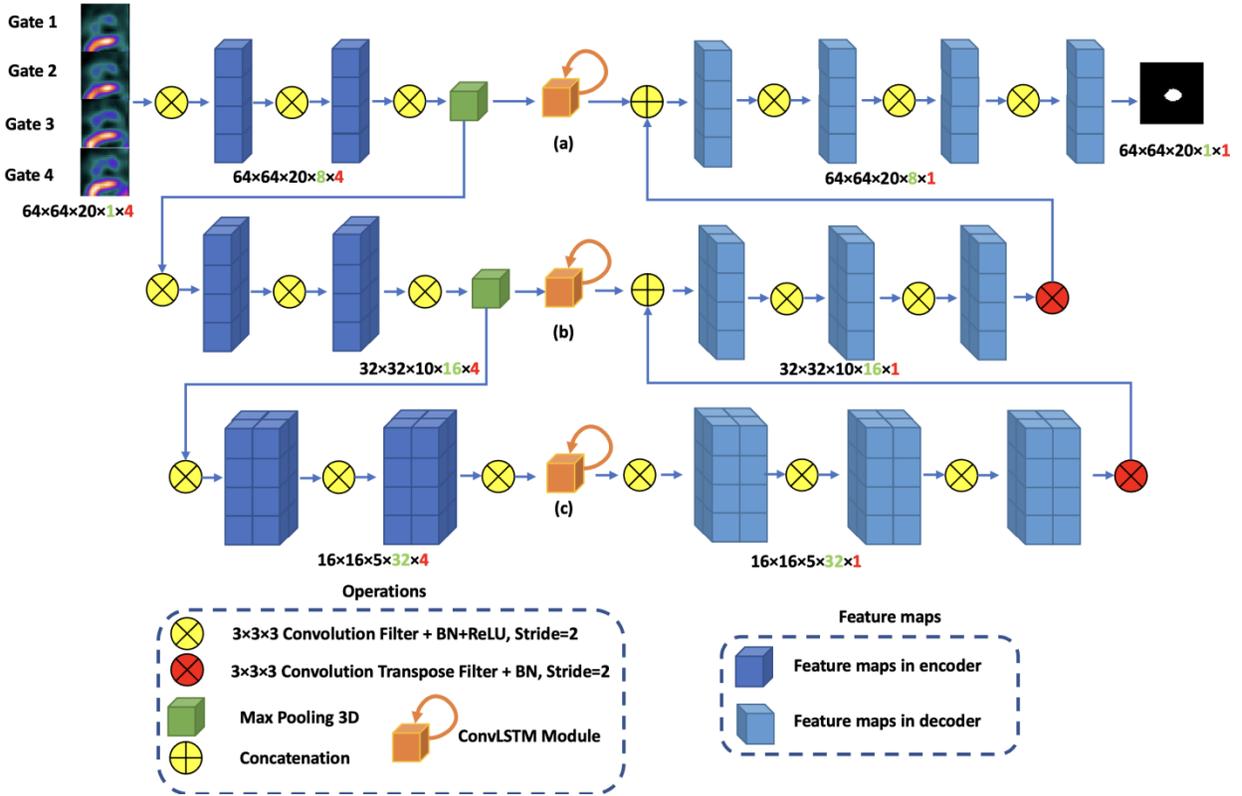

**Figure 1.** The architecture of the proposed ST-VNet for RV segmentation. This figure presents an example using 4 gated MPS as the input and the output is the probability map for the segmentation mask of the 4-th gated MPS. The corresponding black, green and red numbers indicate the dimension of the feature maps, the number of the feature map and the length of the temporal sequence.

1) *V-Net for spatial feature extraction*. V-Net was proposed by [16], which has been widely used for 3D medical image segmentation. The presented V-Net contains an encoder and a decoder. The encoder contains 6 convolutional layers and 2 max-pooling layers to jointly extract hierarchical features from the spatial aspect, which enhances the feature representation ability of the convolutional neural network (CNN) [17]. The decoder contains 10 convolutional layers to restore the down-sampled features and generate the final predictions. The skip connections from the down-sampling path to the up-sampling path are adopted to recover spatially detailed information by reusing feature maps. A concatenation operator is adopted to fuse features from the encoder and decoder. In each CNN layer, a batch normalization layer [18] and a ReLU activation layer are employed to prevent the model from overfitting and accelerate model training.

2) *Convolutional LSTM (ConvLSTM) for sequence modeling*. Fully connected LSTM has proven powerful performance for handling temporal correlations [19]. It contains an input gate, a forget gate and an output gate. The major innovation of LSTM is its memory gate which gradually accumulates the state information along with the input sequence. Every time when the input feature maps $x_t$ come, the memory cell $c_t$ accumulates the information if the input gate $i_t$ is activated. Meanwhile, the past memory $c_{t-1}$ is removed if the forget gate $f_t$ is activated. The output of the LSTM is further controlled by the output gate $o_t$ and the current memory cell $c_t$.

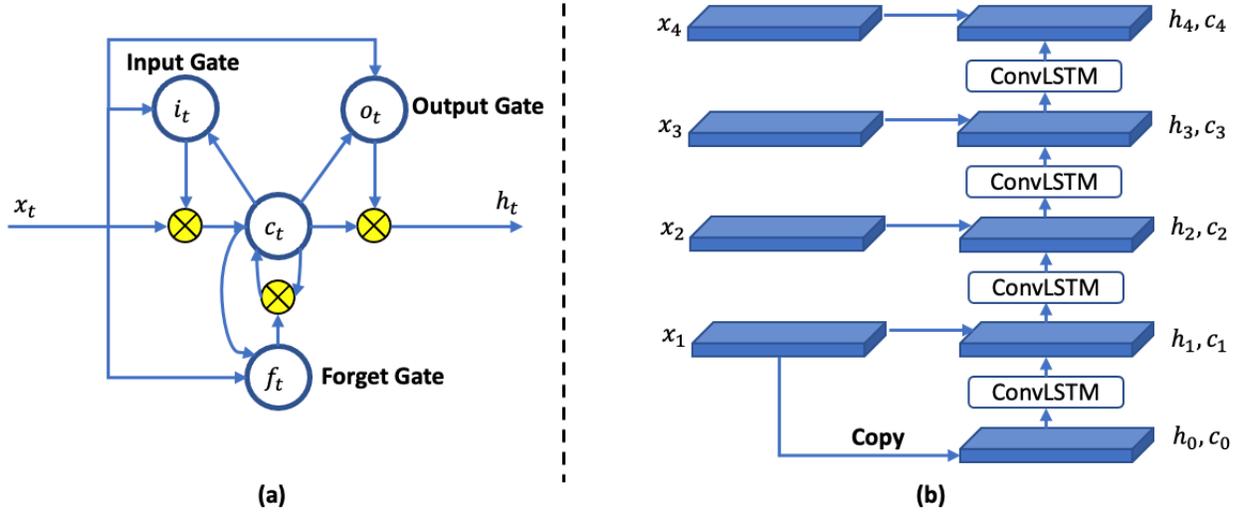

**Figure 2.** (a) Architecture of the ConvLSTM unit. (b) Dataflow of ConvLSTM for processing sequential data with 4 cardiac frames of MPS.

However, the major deficiency of fully connected LSTM in handling spatiotemporal data is its requirement for converting feature maps into feature vectors so that the spatial information will be lost. By replacing the fully connected layers in the LSTM unit with the convolutional operations in the input gate, forget gate, and output gate, the ConvLSTM was proposed [20]. The detailed architecture of an LSTM cell is shown in Figure 2 (a) and the equation of ConvLSTM is shown in Eq. 2.

$$\begin{aligned} i_t &= \sigma(W_i * x_t + U_i * h_{t-1} + V_i \circ c_{t-1}) \\ f_t &= \sigma(W_f * x_t + U_f * h_{t-1} + V_f \circ c_{t-1}) \\ c_t &= f_t \circ c_{t-1} + i_t \circ \sigma(W_c * x_t + U_c * h_{t-1}) \\ o_t &= \sigma(W_o * x_t + U_o * h_{t-1} + V_o \circ c_t) \\ h_t &= o_t \circ \sigma(c_t) \end{aligned} \quad (2)$$

where $*$ represents the convolutional operator in Figure 2 (a) and $\circ$ represents the Hadamard product. The $W$, $U$, and $V$ indicate the convolutional kernels in corresponding gates. $x_t$ is the input of the ConvLSTM module and $h_t$ is the hidden output.

By stacking multiple ConvLSTM layers, we build a network to encode both spatial and temporal features of the MPS sequence. The initial memory cell is copied from the input feature maps and the memory cell is updated according to the input sequential MPS. In detail, the initial hidden state and the memory cell are copied from the feature maps $x_1$ extracted by V-Net from the first MPS $\mathcal{X}_1$. Then the ConvLSTM dynamically accumulates the hidden states and the feature maps from the memory cell to calculate the last hidden state, i.e. $h_4$. The inner dataflow of the ConvLSTM is shown in Figure 2 (b). In our implementation, the ConvLSTM module in Figure 1 contains two ConvLSTM layers, and each gate has 8 feature maps. By applying ConvLSTM layers to the sequential MPS, the feature maps for the last MPS are obtained. Then the feature maps are directly concatenated with the feature maps from the transpose layer in the decoder of V-Net.

### 2.3. Optimization and Evaluation

The proposed RV myocardium segmentation method consists of a training stage and a testing stage. We applied a 5-fold cross-validation to train and test the model. A compound loss function that consists of a DSC loss and an L1 regularizer was employed to optimize the model weights. The DSC measures the overlaying between the segmented volumes and the ground truth, as defined in Eq. 3.

$$DSC = \frac{2|\mathcal{Y} \cap \mathcal{Y}'|}{|\mathcal{Y}| + |\mathcal{Y}'|} \tag{3}$$

where $\mathcal{Y} \in R^{H \times W \times D}$ is the ground truth of epicardium or endocardium, $\mathcal{Y}' \in R^{H \times W \times D}$ is the predicted mask of epicardium or endocardium and $|\cdot|$ indicates the number of foreground (epicardium or endocardium) voxels within the specific volume. To penalize the discrepancy between the generated myocardium mask and the ground truth, and deal with the data imbalance issue [21,22], a DSC loss is adopted to optimize the weights of the ST-VNet. In addition, to prevent model overfitting, L1 regularization is added to the loss function. The overall objective function is defined in Eq. 4.

$$Loss = -\frac{2|\mathcal{Y} \cap \mathcal{Y}'|}{|\mathcal{Y}| + |\mathcal{Y}'|} + |W|_1 \tag{4}$$

where $W$ represents the weights of ST-VNet and $|W|_1$ is the L1 regularizer.

The designed ST-VNet was implemented in Python using Pytorch 1.10 with the backend of CUDA version 11.5. We trained the model on a workstation with an NVIDIA RTX 3090 with 24GB VRAM. Each model was trained for 1000 epochs using an Adam optimizer with a learning rate of 0.001. During the model training, we adopted random flip and random rotation within 15 degrees to perform data augmentation.

To evaluate the model performance, we first checked the generated contours of the epicardium and endocardium visually. To quantitatively reflect the voxel difference, we adopted DSC, sensitivity (SN), and specificity (SP) to evaluate the model performance. The SN measures the proportion of the true positive (TP) voxels compared to the number of TP and false negative (FN) voxels, while the SP measures the proportion of true negative (TN) voxels compared to the number of TN and false positive (FP) voxels. The definition of SN and SP is shown in Eq. 5 and Eq. 6. A DSC, SN, and SP of 1 indicate a perfect match between the model prediction and the ground truth; on the contrary, 0 represents a total mismatch.

$$SN = \frac{TP}{TP + FN} \tag{5}$$

$$SP = \frac{TN}{TN + FP} \tag{6}$$

To measure the surface distance between the predicted masks and the ground truth, we employed Hausdorff distance (HD) and average surface distance (ASD) to evaluate the model. The Hausdorff distance measures the maximum distance between the predicted epicardial/endocardial surface and the ground truth. The ASD indicates the average distance between the predicted epicardial/endocardial surface and the ground truth.

Moreover, we calculated the right ventriclar ejection fraction (RVEF) to further evaluate the proposed ST-VNet. The volume inside the endocardial surface of the myocardium is defined as the RV cavity. The maximum and minimum volumes inside the epicardial surface of the myocardium are denoted as the end-diastolic volume (EDV) and end-systolic volume (ESV), respectively. The RVEF is then calculated as:

$$RVEF = \frac{EDV - ESV}{EDV} X\ 100\% \tag{7}$$

We evaluated the consistency between the RVEF derived by the model segmentation results and the ground truth using mean squared error (MAE), root mean absolute error (RMSE) and Pearson correlation coefficient (PCC). The definition of MAE, RMSE and PCC are shown in Eqs. 8 to 10.

$$MAE = \frac{1}{N} \sum_{1}^{n} |r_e - r_g| \tag{8}$$

$$RMSE = \frac{1}{N}\sum_{1}^{n}\sqrt[2]{(y_i - \hat{y}_i)^2} \qquad (9)$$

$$PCC = \frac{\sum_{i=1}^{n}(r_e^i - \overline{r_e})(r_g^i - \overline{r_g})}{\sqrt{\sum_{i=1}^{n}(r_e^i - \overline{r_e})^2}\sqrt{\sum_{i=1}^{n}(r_g^i - \overline{r_g})^2}} \qquad (10)$$

where $r_e^i$ indicates the measured RVEF using segmentation results, and $r_g^i$ represented the ground truth of the RVEF. $N$ is the sample size and $i$ is the index of the training sample. And $\overline{r_e}$ and $\overline{r_g}$ are the mean values of the corresponding measurements.

### 3. Experimental Results

#### 3.1. Baseline Models

To reflect the effectiveness of the proposed model, two baseline models were implemented by removing the ConvLSTM module and V-Net module, thus a plain V-Net and a Spatial-Temporal-Net (S-T-N) were obtained. The architectures of the baseline models are shown in Figure 3.

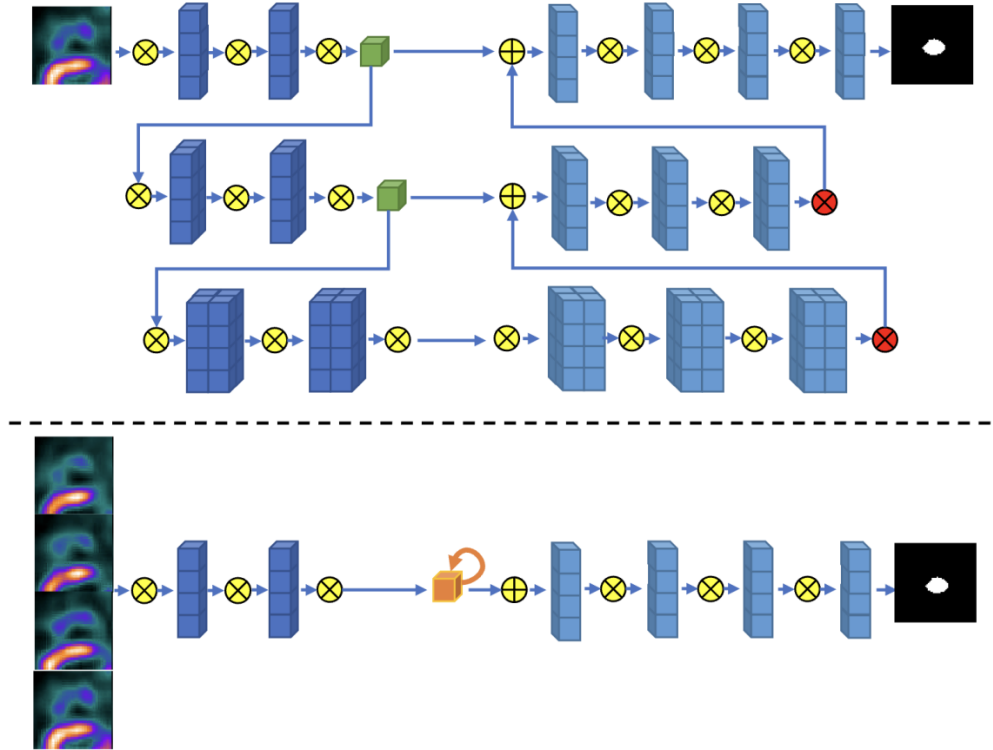

**Figure 3.** Baseline models. *Top*: V-Net, which is obtained by removing the ConvLSTM modules and its input contains only one gate (cardiac frame). *Bottom*: ST-Net, which is obtained by removing the pooling layers in V-Net.

#### 3.2. Results of Myocardium Segmentation

We tested the models with different gates of MPS as the sequential MPS, and Table 1 quantitatively compared the model performance of the baseline models and the proposed ST-VNet for RV epicardium and endocardium segmentation, respectively. The performance was reported according to the subjects from the test set in each cross-validation fold.

**Table 1.** Quantitative evaluation for the V-Net, ST-Net, and ST-VNet for RV epicardium and endocardium segmentation. DSC, Dice similarity coefficient; HD, Hausdorff distance in pixel; ASD, average surface distance in pixel; SN, sensitivity; SP, specificity. The bold text indicates that the model achieved the best performance within the corresponding task.

| Class | Model | Number of gates | DSC | HD (pixel) | ASD (pixel) | SN | SP |
|---|---|---|---|---|---|---|---|
| Epicardium | V-Net | 1 | 0.8085±0.0340 | 7.1753±1.0232 | 3.1068±0.8381 | 0.8268±0.0492 | 0.9374±0.0195 |
| | ST-Net | 3 | 0.8685±0.0380 | 6.9836±1.7825 | 1.8490±0.9061 | 0.8769±0.0402 | 0.9574±0.0139 |
| | ST-VNet | **2** | **0.8914±0.0311** | **6.0288±2.2818** | **1.3479±0.7355** | **0.8993±0.0381** | **0.9638±0.0124** |
| Endocardium | V-Net | 1 | 0.5717±0.0923 | 7.4620±1.6624 | 7.6182±2.7494 | 0.6722±0.0976 | 0.9735±0.0088 |
| | ST-Net | 6 | 0.7986±0.0662 | 6.3434±2.8989 | 3.2448±2.6173 | 0.8151±0.0788 | 0.9916±0.0038 |
| | ST-VNet | **3** | **0.8157±0.0624** | **4.9917±2.3557** | **2.5411±2.1086** | **0.8230±0.0802** | **0.9924±0.0035** |

According to Table 1, the proposed ST-VNet achieved the highest DSC, SP, and SN, and the lowest HD and ASD on our MPS dataset for both RV epicardium and endocardium segmentation. The epicardial and endocardial contours generated by different models of the representative slices are illustrated in Figure 4.

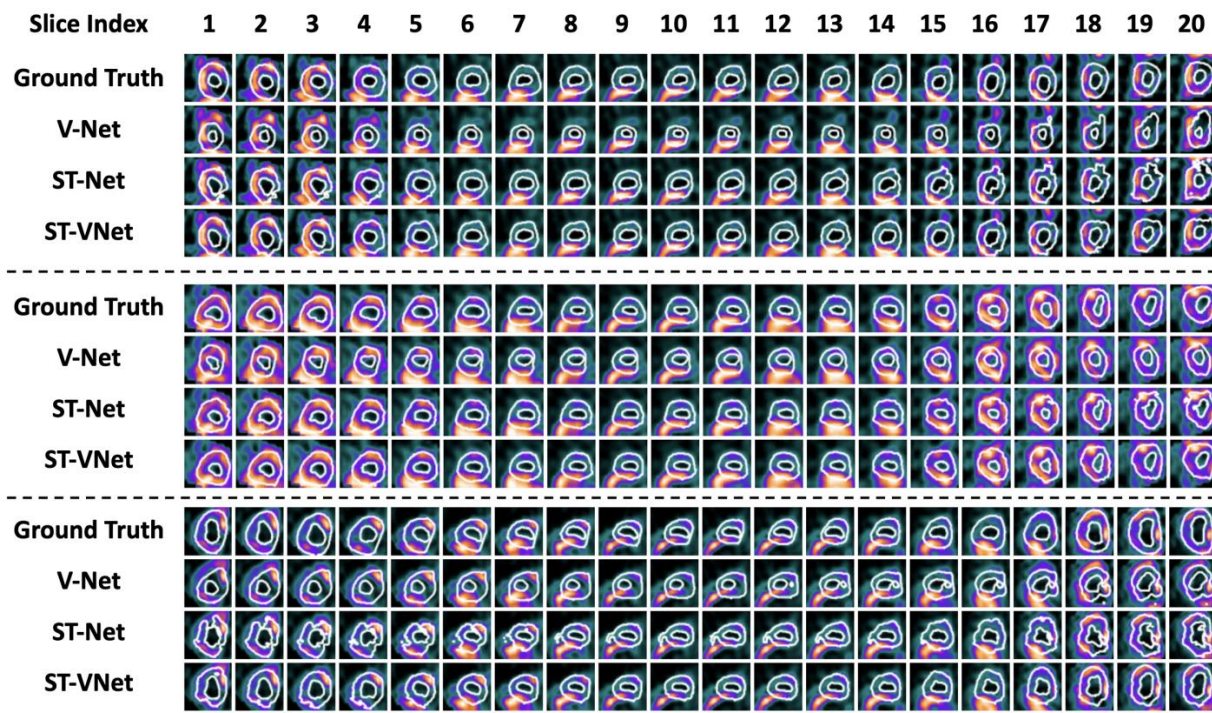

**Figure 4.** Comparison of the segmentation results for different models in axial view. *Top*: patient #13 gate #7; *middle*: patient #17 gate #6; *bottom*: patient # 33 gate #5. For each grouped image, the epicardial and endocardial contours generated by manual annotation (Ground Truth), V-Net, ST-Net, and ST-VNet are depicted.

We further validated the performance of the proposed ST-VNet and baseline model ST-Net using MPS sequence with different numbers of gates, as shown in Figure 5.

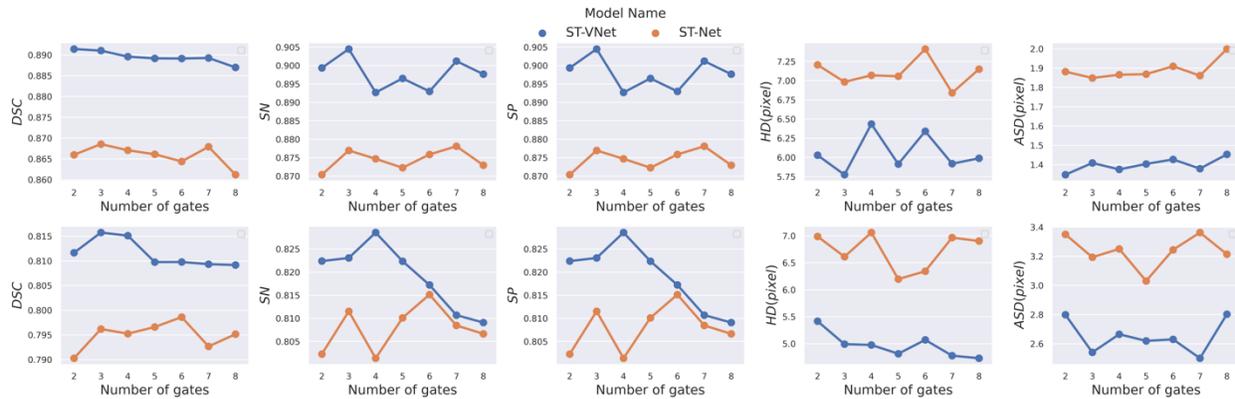

**Figure 5**. Quantitative evaluation of ST-VNet and ST-Net for RV epicardium and endocardium segmentation. *Top*: epicardium segmentation; *bottom*: endocardium segmentation. The horizontal axis indicates the number of MPS gates. DSC, Dice similarity coefficient; HD, Hausdorff distance in pixel; ASD, average surface distance in pixel; SN, sensitivity; SP, specificity.

According to Figure 5, the proposed ST-VNet achieved the best DSC using the MPS sequence with 2 and 3 gates for epicardium and endocardium segmentation. In addition, the ST-VNet consistently outperformed ST-Net using MPS sequence with different numbers of gates in all evaluation metrics.

For each subject, 8 gated MPS images were scanned. We further evaluated the segmentation performance for each MPS gate using V-Net, ST-Net, and ST-VNet. The results are shown in Figure 6.

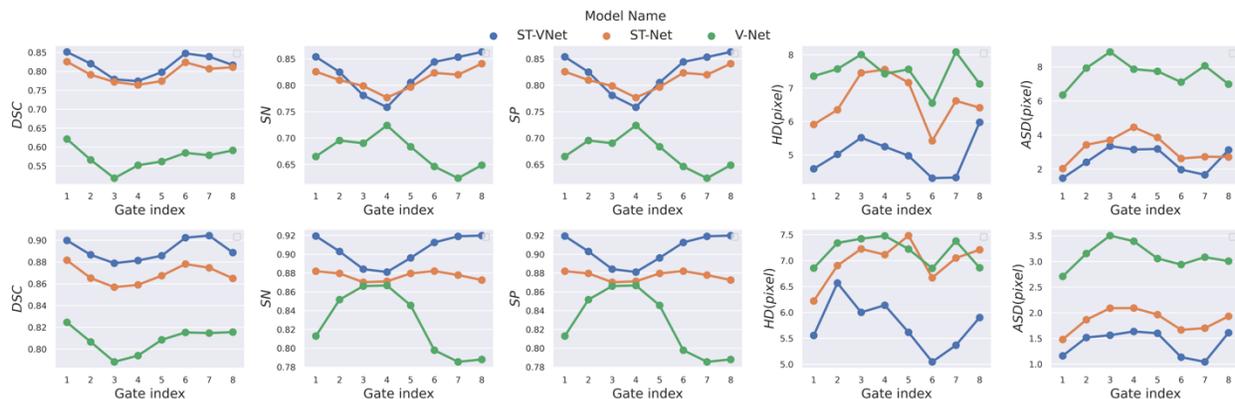

**Figure 6**. Quantitative evaluation of RV epicardium and endocardium segmentation for each MPS gate. *Top*: segmentation; *bottom:* endocardium segmentation. DSC, Dice similarity coefficient; HD, Hausdorff distance in pixel; ASD, average surface distance in pixel; SN, sensitivity; SP, specificity.

### 3.3. Results of RVEF quantification

We firstly inspected the volumes of the segmentation results and compared the correlation of the RV myocardium volumes between the ground truth and those calculated using the predicted contours. The calculated RV myocardium volumes for each gate are shown in Figure 7.

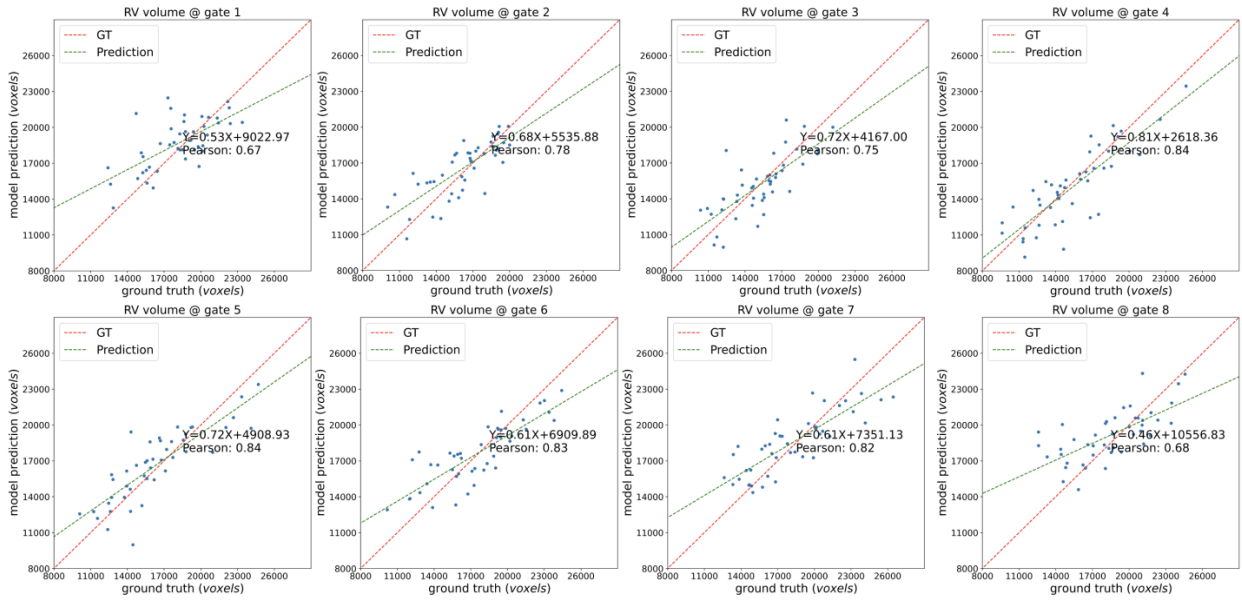

**Figure** 7. RV myocardium volumes between ground truth and proposed method for each gate among 45 subjects. Each blue dot indicates the RV myocardium volume of a subject. The red dashed line represents the identity line, and the green line indicates the linear regression results.

We calculated the RVEF using the segmented RV endocardium and compared it with the ground truth RVEF using the manually annotated RV endocardium. The results are shown in Figure 8.

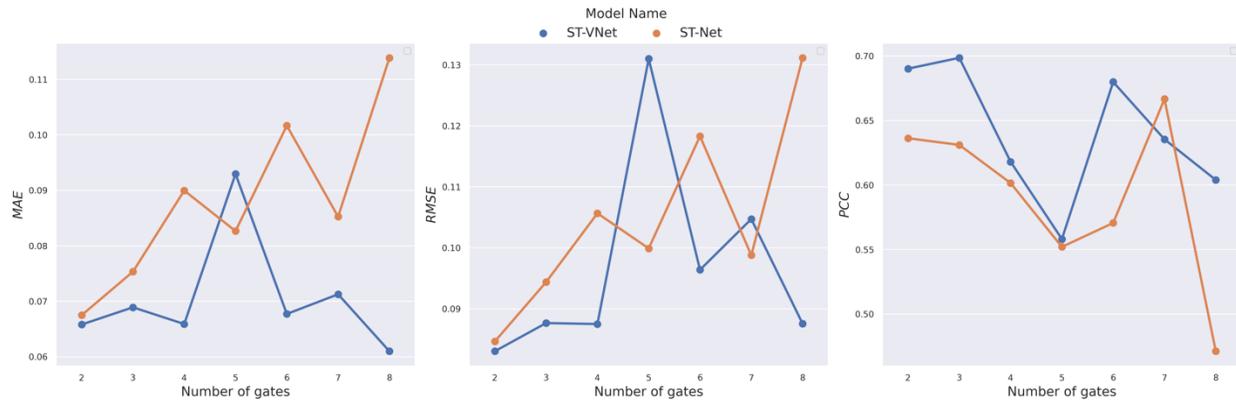

**Figure 8**. Evaluation of RVEF quantification for ST-Net and ST-VNet trained with MPS sequence with different numbers of gates. MAE: mean squared error; RMSE: root mean absolute error; PCC, Pearson correlation.

## 4. Discussion

### 4.1. Performance of RV segmentation

In this study, the results strongly demonstrate the feasibility of the proposed model in clinical practice for the delineation of the RV epicardial and endocardial contours. According to Table 1, the proposed ST-VNet achieved the highest DSC, SN and SP, and the lowest HD and ASD for epicardium and endocardium segmentation compared to those two baseline models. For epicardium segmentation, the ST-VNet improved the DSC from 0.8085 achieved by V-Net to 0.8914, and decreased ASD from 3.1068 pixels achieved by V-Net to 1.3479 pixels; for endocardium segmentation, the ST-VNet improved the DSC from 0.5717 achieved by V-Net to 0.8157, and decreased ASD from 7.6182 pixels achieved by V-Net to 2.5411 pixels.

In Figure 4, the comparison of the segmentation results indicated that the epicardial and endocardial contours generated by ST-VNet were closer to the ground truth than those of V-Net and ST-VNet among the three representative subjects. The comparison between the ST-VNet and V-Net indicated that adding ConvLSTM layers to

V-Net was effective for capturing temporal features according to MPS sequence, and thus the ST-VNet achieved better performance. For example, for patient #33 at the bottom of Figure 4, the predicted contours of the ST-VNet in slices 1 to 3 and slices 17 to 20 were smoother than those of V-Net. The comparison between the ST-VNet and ST-Net demonstrated that adding pooling layers to hierarchically extract spatial features was important for RV segmentation. For example, the endocardial contours in slices 16 to 20 among these three subjects generated by ST-Net were rough, sharp and uneven; however, the ST-VNet generated contours were smooth and close to the ground truth. The results showed that the ST-VNet was effective in modeling spatiotemporal data for RV myocardium segmentation using 4D images.

From Figure 5, we can observe that for epicardium segmentation, when the number of gates of the MPS was set as 2, the ST-VNet achieved its best performance; however, for endocardium segmentation, with the increase of gates in the MPS sequence, the performance of ST-VNet firstly increased to its peak when the number of gates of the MPS was set as 3, and then slightly degraded. This was because integrating contextual temporal features was important for video prediction [23], and the cardiac motion was traceable within a cardiac cycle. However, increasing the gates in a sequence greedily, i.e. from 3 to 8, didn't further improve the segmentation performance.

For each gate, we compared the segmentation performance between the proposed ST-VNet and those two baseline models in Figure 6. The visualization results indicated that the ST-VNet trained using MPS sequence with two gates outperformed ST-Net in DSC for epicardium segmentation among different gates, and consistently outperformed V-Net and ST-Net in all evaluation metrics for endocardium segmentation among different gates. However, for the third gate and the fourth gate, the ST-Net slightly outperformed the proposed ST-VNet in SN and SP; for the the eighth gate, the ST-Net achieved a lower ASD than ST-VNet. Compared to V-Net, the proposed ST-VNet improved the segmentation performance significantly in all evaluation metrics among different gates. These findings demonstrated the superior performance of the proposed ST-VNet.

**4.2. Performance of RVEF quantification**

According to Figure 7, the predicted volumes and ground truth showed a high Pearson correlation using the proposed ST-VNet. For gates of 4 to 7, the Pearson correlation was higher than 0.82. According to Figure 8, the proposed ST-VNet achieved the lowest MAE and RMSE of 0.0609 and 0.0830 for RVEF calculation using MPS sequence with 8 and 2 continuous gates. Furthermore, the ST-VNet achieved its highest Pearson correlation of 0.6985 using MPS sequence with 3 gates. The precise segmentation model improved RVEF quantification.

**4.3. Clinical values of RV segmentation and functional assessment**

During the past decades, RV functional assessment was neglected in the diagnosis and prognosis of left-sided heart failure. However, RV dysfunction has been left-sided heart failure with both reduced ejection fraction (HFrEF) and preserved ejection fraction (HFpEF), and strongly contributed to the increase of morbidity and mortality [24,25]. Right heart function is an indicator to measure the severity of pulmonary hypertension; pulmonary arterial hypertension (PAH) often becomes increasingly complicated due to right heart dysfunction. Moreover, improvement of right heart function is important to evaluate the therapeutic effect and prognosis of PAH. Therefore, it is essential to measure valuable parameters to assess RV function in HF and PAH patients. Accurate assessment of RV function is also likely to guide implantation of ventricular assist device (VAD) among those with medically refractory right sided heart failure.

Right heart catheterization (RHC) is commonly used to measure RV hemodynamic parameters and is the gold standard for PAH diagnosis. However, first, RV cardiac output (CO) is commonly measured by thermodilution in RHC, which may overestimate CO in patients with decreased cardiac function [26]. Second, as an invasive test, RHC requires hospitalization, which limits its clinical application in primary diagnosis and follow-up. Echocardiography is the most widely used right heart functional test, but it also has many disadvantages: (1) there is no echocardiographic gold standard of RV functional assessment; (2) it has poor reproducibility; (3) two-dimensional echocardiographic measurements are obtained in a four-chamber view, which is only a single view. MPS has been widely used in the diagnosis of coronary heart disease and HF. If the indicators measuring RV function in MPS were discovered, it would benefit a large number of patients by reducing patient examinations and saving medical resources.

## 5. Conclusion

We proposed a spatial-temporal V-Net to automatically extract RV epicardial and endocardial contours from gated MPS. The proposed ST-VNet fully used the spatial features from the 3D gated MPS by V-Net, and the temporal correlations from the MPS sequential imaging using ConvLSTM. It was validated on our dataset with 45 subjects and produced accurate results for both RV segmentation and RVEF measurement. The proposed deep learning model would provide auxiliary support in clinical practice for RV functional assessment in MPS.

## Acknowledgment

This research was supported by a new faculty startup grant from Michigan Technological University Institute of Computing and Cybersystems (PI: Weihua Zhou) and a seed grant from Michigan Technological University Health Research Institute (PI: Weihua Zhou).